\documentstyle[eqsecnum,aps,prd]{revtex}
\def\part{\partial}

\def\be{\begin{equation}}
\def\ee{\end{equation}}
\def\eq{\be}
\def\en{\ee}
\def\dtwo{d\Omega_{(2)}}

\begin{document}

\begin{titlepage}
\begin{center}
\today     \hfill    LBNL-38560\\
{}~           \hfill UCB-PTH-96/10\\
{}~           \hfill astro-ph/9605132 \\

\vskip.7in {\Large\bf
Open Universes from Finite Radius Bubbles}
\vskip.3in
J. D. Cohn\footnote{jdc@asterix.lbl.gov}\\
{\it Department of Physics, University of California, and\\
Theoretical Physics Group,
Lawrence Berkeley National Laboratory\\
University of California
Berkeley, CA 94720}
\end{center}

\vskip.2in
\begin{abstract}
The interior of a vacuum bubble in de Sitter space
may give an open universe with sufficient homogeneity to
agree with observations.  Here, previous work by
Bucher, Goldhaber and Turok is extended to describe a thin bubble
wall with nonzero radius and energy difference across the wall.
The vacuum modes present before formation of the bubble
propagate into the interior of the open universe and 
the power spectrum of the resulting gauge
invariant gravitational
potential is calculated.  It appears to become scale invariant
on small scales,
with onset at about the same scale
as that found in the zero radius case.  There is sensitivity
to the radius and energy difference at large scales, but it is 
expected that they 
cannot be strongly constrained because of cosmic
variance.  The prediction of a scale invariant spectrum
seems to be robust with respect to variation of these parameters at
small scales, and apparently is a
generic feature of the contribution of these modes 
for these thin wall models.
\end{abstract}

\begin{center}
PACS numbers 98.80Bp, 98.80Cq
\end{center}

\end{titlepage}
\setcounter{footnote}{0}

\section{Introduction and background}
Inflationary models usually predict $\Omega = 1$ \cite{guth}, but
it has recently been shown they
also allow for the possibility of
 open\cite{BGT,cmb,amen,linde,css} and
closed\cite{lm} universes.  The current models of open universes
are either based on nucleation of vacuum
bubbles\cite{gott,col1} or
on open universes which are also compact\cite{css}.
This paper extends work on models based
on bubble nucleation.

Bubble nucleation occurs when a field configuration
trapped in a false vacuum tunnels through a barrier
to the true vacuum.  The tunneling only happens
over a finite region of space, so that a bubble of
true vacuum forms, with a bubble wall (a domain wall)
interpolating between the false vacuum
outside and the true vacuum inside.  The thickness of
the bubble wall and the size of the bubble are determined by
parameters in the theory.
If there is more than one field,
the bubble wall interpolates between true and false vacuum
values of all the fields in the theory.

The interior of a vacuum bubble
is an open universe\cite{col1,gott}.
If there are random initial conditions before nucleation,
the open universe in the bubble interior
has primordial inhomogeneities, which may be
erased by a subsequent phase
of inflation.
However, the amount of inflation required to
obtain sufficient homogeneity to agree with observation
will also make the universe
approximately flat, with a critical density $\Omega \approx 1$.
Recently proposed\cite{BGT,cmb,linde,amen}
bubble models accomplish homogeneity without flatness by having
three stages.  First there is a phase of
ordinary inflation, driving
$\Omega \rightarrow 1$ and erasing initial
inhomogeneities.
Then a bubble is nucleated, resetting
 $\Omega \sim 0$ in the interior of the bubble.
A period of slow roll inflation then follows, solely for the
purpose of driving
$\Omega$ up to some desired value slightly less than one.
At the
end of this second stage of inflation, the usual inflationary
scenario for reheating takes over.
The open universe is the interior of the forward light cone of the
center of the bubble.  The bubble wall, exterior to this light cone
by definition\footnote{although for a thick wall there may
be some overlap}, has moved out to infinity
before some time $T$
in the past.

Modifications using two fields
\cite{linde,amen,lm} can result in universes with
variable values of $\Omega$.
One field provides inflation and the other field resets $\Omega
\sim 0$ through bubble nucleation
 at some point.  The advantage of this is that more
natural (e.g.  polynomial) potentials can be used for the fields.
(The one field models need more complicated potentials to
resemble the calculable toy models, although these might
occur in a supergravity model.)
Some care must be taken to avoid large inhomogeneities
due to the change of natural time coordinate after tunneling
(the synchronicity problem).  In addition, predictability can be
limited because the amount of inflation, depending on one
of the fields, may
not necessarily
be correlated with the random tunneling time of the other field,
producing a range of different $\Omega$'s.  However,
there are modifications which can give more
 predictability\cite{linde,amen}.  As mentioned in
\cite{BGT}, much of the one field model analysis carries over to
two field models.

The three stage scenario has some general properties \cite{BGT}:
Before the bubble nucleates, the first stage of inflation
drives spacetime to
the de Sitter invariant vacuum, the Bunch-Davies
vacuum.
The Bunch-Davies vacuum was argued
in \cite{BGT} to be the
natural vacuum because it has the right symmetries and
it goes over to the
Minkowski vacuum at short distances.
  This
provides some robustness with respect to initial conditions.
The recurrence of inflation after tunneling
means the bubble must tunnel from
de Sitter space to de Sitter space.
The slow roll in the second phase of inflation means that the
effective mass after tunneling should be approximately $m^2 \sim 0$.
The second stage of inflation must be fine tuned
({\it i.e.} the parameters in the potential must be fine tuned)
within a few percent in order to get $\Omega \sim .1 - .9$.

It is of interest to know how the particulars of the bubble formation
affect the resulting density perturbations.  Unlike ordinary
inflation, here some fine tuning of the parameters in the potential
is required to fix $\Omega$, so it is possible that other properties
of the resulting theory are sensitive to variation of potential
parameters.
The focus here is on the imprint
of the bubble wall on
the vacuum density perturbations.
There are also perturbations arising from
the fluctuations of the bubble wall itself
\cite{linde,lm,juan,HSTY,gar96}
which must be included.

In order to calculate the density perturbations one does the following:
The density perturbations start out (since
an initial long period of inflation is assumed) as
Bunch-Davies vacuum modes of the false vacuum corresponding to a
field of some mass $M$.
Tunneling is formally a process in Euclidean time, so
these modes are analytically continued to Euclidean time to give
initial conditions.
The bubble wall is a boundary between two de Sitter spaces,
which will be called the false and true vacua from here on
(the final space is not actually a vacuum but
 the slow rolling second stage of
inflation, it will be called the true vacuum here since it
is the endpoint of the tunneling).
The initial false vacuum fluctuations propagate through
the bubble wall and are matched onto a configuration of
true vacuum modes
using Bogoliubov coefficients.
At the nucleation time,
the whole system is analytically continued to Lorentzian
signature.
 Using the natural coordinate system for matching
across the wall means that the resulting wavefuction must
be normalized by hand at some point and can be done here.
The perturbations, once inside the forward light
cone of the bubble (which requires another analytic continuation,
since the natural
coordinate system is singular on the light cone), are
related to perturbations of the gauge invariant gravitational
potential, $\Phi$.
The power spectrum of $\Phi$, eq. (\ref{powergen}) along with
eqns.(\ref{alphabog},\ref{betabog}), is the main result
of this paper.
It can be combined with contributions from the wall fluctuations
and some model of matter content in
the universe ({\it e.g.}, cold dark matter) to get
temperature fluctuations to compare
to the CMB.

The formalism for fluctuations around a
tunneling background has been developed in many papers, see
\cite{wkb-tun,rub,kandrup,vv,wkb-kyot,TS} and references therein.
Bogoliubov coefficients for specific thin and thick bubble walls (both
designed to be tractable and yet have enough parameters to
be able to test for robustness), were
calculated in Minkowski space in \cite{yts,HSTY}.
In these the fluctuations were massive in both the true and the false
vacuum.

The inclusion of gravity and the true vacuum mass $m^2 \approx 0$
add extra complications.
In \cite{BGT}, the wall
was taken to be a step function, and the
radius of the bubble at nucleation, $R$, was taken to be zero.
The power spectrum was calculated
for a field with initial false vacuum mass
$M^2=2H^2$, and the metric perturbations corresponding to the field
fluctuations were identified.  They also found the matching conditions
for a massless field across the light cone, which is subtle due
to only half of the modes of a massless field propagating
into the light cone.
A second paper, \cite{BT}, generalized this to a field with arbitrary
mass in the false vacuum, but still with radius zero.
In \cite{TS}, the quantum
state was found for a thin wall bubble with gravity when the fluctuations
only have mass on the wall itself.

Effects of perturbations of the wall itself have also
been studied in \cite{lm,juan,gar96}, building on
work on bubble fluctuations by \cite{gv}.
The temperature fluctuations predicted in the CMB from
Bunch-Davies vacuum fluctuations (without a bubble) have been
calculated in \cite{BY} and compared to the CMB measurements
and predictions of the conformal \cite{conformal} vacuum.

In this paper, the thin wall approximation is retained, but
the bubble wall is given a finite radius $R$, for an initial false
vacuum mass
conformal or larger, $M^2 \ge 2 H^2$.  Some of the other assumptions
used previously are also carried over: that the
era preceeding tunneling has resulted in a homogeneous Bunch-Davies
vacuum as initial conditions and that
bubble nucleation is rare enough that
no other bubbles are nearby.
First the background solutions, approximations and fluctuation modes
are reviewed.  After finding Bogoliubov coefficients, the wavefunction
is normalized and continued into the forward light cone of the
bubble interior (the open universe).
Then the power spectrum for the
gauge invariant gravitational potential is calculated for this
specific background using the results of \cite{BGT} and compared to
earlier cases.
These results must be combined with those for fluctuations of the
wall itself and some model for structure formation ({\it e.g.},
cold dark matter) in order to make a prediction for experiments.

Note Added:  While this paper was being prepared for submission,
the preprint\cite{new} appeared, which independently considers
some of the same issues.  Some puzzles pointed out in this paper are 
addressed therein; the final version of this paper has been
updated to refer to these where appropriate.

\section{Background solution, coordinate systems, and approximations}
The background bubble solution for the fluctuations is reviewed
first\cite{col1,rivers,TS}.
The bubble is described to first approximation by classical
evolution in Euclidean space.  After nucleation, the system is
rotated to Lorentzian signature and continues evolving
classically\cite{col1,rivers}.

The action for a single scalar field $\phi$ plus gravity is
\be
S = \int d^4 x \, \sqrt{-g}[{1\over 2}g^{\mu \nu}\part_\mu \phi
\part_\nu \phi - V(\phi) - (16 \pi G)^{-1} R]
\ee
with metric $g_{\mu \nu}$, curvature scalar $R$, and
potential $V(\phi)$.
In the flat space case, a solution respecting SO(4) symmetry has
minimum Euclidean action, so when gravity is included,
one looks for a solution depending only
upon the rotationally invariant radial coordinate
$\sigma$.  This has not been proven to be a property of
the minimum action solution for the
case with gravity, but none with smaller action are known.

With gravity, the
Euclideanized metric for tunneling is written in terms of
$\sigma^2$,
\be
ds^2 = d \sigma^2 +b^2(\sigma)[d\tau_E^2 + \cos^2\tau_E \dtwo]
 \; \; ,
\ee
with $ \dtwo = d \theta^2 + \sin^2 \theta d\phi^2 $.

The Euclidean equations of motion are then
\be
\phi''(\sigma) + 3{b'(\sigma)\over b(\sigma)} \phi'(\sigma) =
{\part V \over \part \phi} \label{euceom1}
\ee
and
\be
\left( {b'(\sigma ) \over b(\sigma ) }\right)^2 = {1 \over b^2(\sigma)}
+ {8 \pi G \over 3}\left[{1\over 2} \phi'^{ \; 2} - V[\phi(\sigma)] \right] \;
{}.
\label{euceom2}
\ee

The boundary conditions are
\be
\begin{array}{l}
b(\sigma = 0) = b(\sigma = \sigma_{max}) = 0 \; , \\
\phi'(\sigma = 0) = \phi'(\sigma=\sigma_{max}) = 0 .
\end{array}
\ee
where the turning point in the tunneling solution,
$\sigma_{max}$, is infinite for the flat space case,
but becomes finite once gravity is included\cite{hawkmoss,stein}.
For $\phi \approx$ constant,
define  $H^2 = 8 \pi G V(\phi) / 3$,
so that for de Sitter space, $b(\sigma) = H^{-1}\sin[H\sigma]$.
For flat space, $b(\sigma) = \sigma$.

For many of the calculations it will be useful to switch to
a second coordinate system, defined by
$\tanh u = \cos[H\sigma]$.
In terms of these coordinates, the Euclidean line element is:
\be
ds^2 = a^2(u)(du^2 +
d\tau_E ^2 + \cos^2\tau_E \dtwo) \; ,
\ee
for de Sitter space,
\be
a^2(u) =
 {1 \over H^2 \cosh^2 u} \; .
\ee
As one approaches the light
cone, $\sigma \rightarrow 0$, and $ u \rightarrow \infty$.

The equations of motion (\ref{euceom1},\ref{euceom2})
have not been solved in general.
A few different approximations are possible, whose validity depends on
parameters in $ V(\phi)$.
A standard approximation is the thin wall approximation, where
the energy difference between the true and
false vacuum, $\epsilon^2 =  H_F^{2}- H_T^{2}$,
is very small\cite{col1}.
The bubble wall, the boundary between the
two vacua, is then located approximately at one value of
$b(\sigma) = R$.
In order to consistently neglect $r_b$, the width of the wall,
it must be small enough compared to other
parameters in the theory,
\eq
r_b \ll R \; , \; \; r_b \ll \epsilon^{-1} \; , \; \;
r_b \ll  (H_F^{2} + H_T^{2})^{-1/2}\equiv \Lambda^{-1} \; .
\label{thinw}
\en
In order to verify these conditions, the radius $R$ in the
presence of gravity can be calculated using the radius $R$
in the absence of gravity,
$\Lambda$, and $\epsilon$ \cite{parke}.

Another possible approximation is to neglect gravity.  The
validity of this is measured by the second term in
equation (\ref{euceom2}), since for flat Minkowski space
$b'(\sigma) = 1$.

In \cite{BGT} a `new thin wall' approximation was made.
They assumed that $R<H^{-1}$, the wall is thin,
the potential is slowly varying after nucleation ($M_{true} \approx 0$),
and that the bubble nucleates at about the same energy as that of the
false vacuum ($\epsilon \approx 0$).
This allows one to treat the background as fixed and
to neglect gravity, which they verified is consistent.
The energies of the initial and final states are
approximately degenerate, $H_F \approx H_T$,
and $R \approx 0$.
After tunneling, the potential is flat enough
for slow roll inflation, so that it is flat enough
for the field $\phi$ to be considered constant.  Thus expressions
for tunneling from de Sitter space to de Sitter space can be used.

The background for fluctuations in this paper is
the thin wall solution found by \cite{TS}, which, in terms of
the coordinate $u$ above, is
\be
a(u) = \left \{ \begin{array}{lr}
        {1 \over H_F \cosh u} \; \; \;& u < u_R \\
        {1 \over H_F[\cosh(u-u_R) \cosh u_R
+ \sinh(u-u_R)\sqrt{\cosh^2u_R - (H_T/H_F)^2}]} & \; \; u >u_R
       \end{array}
\right.
 \label{inst}
\ee
\be
\phi_{b}(u) =
\left\{
\begin{array}{lr}
              \phi_{i} &{\rm if} \; u < u_R \\
              \phi_{n} &{\rm if} \; u > u_R
              \end{array}
\right.
\ee
This solution requires $u_R>0$.
The interior
$a(u)$ can be recognized as $(H_T \cosh(u+\delta))^{-1}$,
the metric for a shifted value of $u$, so that the metrics
are matched at the same value of $a(u)$, which is the radius
\be
R = {1 \over H_F \cosh u_R} = {1 \over H_T \cosh(u_R + \delta)} \; .
\label{radius}
\ee

The coordinate shift $\delta$ obeys
\be
e^{\delta} = {(1-\sqrt{1-(H_FR)^2})
(1+ \sqrt{1-(H_TR)^2}) \over (H_TR)(H_FR)}
\ee
Setting the $H_T = H_F$ takes $\delta \rightarrow 0$.
Taking the matching radius $R \rightarrow 0$ means
(from eq. (\ref{radius}))
$H_T \rightarrow H_F$ and $u_R \rightarrow \infty$.

Once the bubble is nucleated, one rotates to a Lorentzian
signature to continue the evolution.
Usually the nucleation time is chosen at $\tau_E= 0$.
The coordinate system for the Euclidean description of the bubble can
be analytically continued,
\be
\begin{array}{ll}
ds^2  &= d \sigma^2 +b^2(\sigma)[-d\tau^2 + \cosh^2\tau \dtwo] \\
&= {1 \over H^2 \cosh^2 u} [du^2 -d\tau^2 + \cosh^2\tau \dtwo ]
\end{array}
\label{desitmet}
\ee
The equations of motion for the
$\sigma$ dependent background remain unchanged.

On the light cone of the bubble center, $b(\sigma) = 0$ ($u=\infty$),
so there is a coordinate singularity.  One can continue to
the interior of the light cone by rotating,
$\sigma = - i T, \tau = \chi - i \pi/2, b = - i c$, to the
coordinate system with line element
\be
ds^2 = - dT^2 + c^2(T)[d\chi^2 + \sinh^2\chi \dtwo] \; .
\ee
The equations of motion are now
\be
\begin{array}{l}
\ddot{c}(T) + 3{\dot{c}(T)\over c(T)} \dot{\phi}(T) =
-{\part V \over \part \phi} \\
({\dot{c} (T ) \over c(T) })^2 = {1 \over c^2(T)}
+ {8 \pi G \over 3}[{1\over 2}\dot{ \phi}^2 + V[\phi(T)] ] \; .
\end{array}
\ee
For de Sitter space, the background solution except at
the wall (which is not inside the interior of the light
cone for a thin wall) is $c(T) = H^{-1}\sinh[HT]$.

Classically, then, the following picture arises \cite{col1}:
there are two regions of space considered here,
outside the light cone of the
bubble center (which will be called region II, as was done in
\cite{BGT}) and inside the light cone of the bubble center
(region I).\footnote{To cover de Sitter space requires another
region antipodal to I, but it will only be relevant here when
verifying the wavefunction normalization inside region I, summarized
in section 5.}
For the thin and new thin wall approximation,
the bubble wall is approximately
located at just one value of $b(\sigma) = R$.
The wall moves exterior to the light
cone of the bubble's center, approaching it
on the curve $b(\sigma)^2 = R^2$.  (For $R=0$ the bubble wall
is the light cone.)  The bubble wall has no intersection
with any of the curves
in
region I.  There may be some overlap with region I for a wall
that isn't thin.  The energy difference between the true and false
vacuum
goes into accelerating the wall (which as it increases in size,
due to surface tension, contains more energy).

In region I, inside the future light cone,
the SO(3,1) invariant line
element $T^2$ behaves just like time in the equations of motion.
The field $\phi$ is constant on lines of constant $T$ and thus
region I can be viewed as an open  Freedman-Robertson
Walker universe with scale factor $c(T)$.
The constant $T$ surfaces are not Cauchy surfaces, however,
which will be important later on when the spectrum is
considered.  Since a thin bubble wall is confined to region II,
it occurs ``before'' $T=0$ for the open universe.

There are consistency requirements for parameters in
a given background theory, some of which are automatically
satisfied in the thin wall approximation.
The action of the bubble should be large in order for the
semiclassical approximation to be valid, and can
be calculated directly from the action in the absence
of gravity\cite{parke}.  A large action
makes the nucleation rate small, another requirement
for the scenario here,
so that tunneling occurs only after
inflation has driven the initial conditions
into the Bunch-Davies vacuum.
The surface tension controls how much the bubble
wall remains close to its classical configuration, and
thus it must be large for the classical solution to
be a good saddle point.  Increasing the surface tension increases
the action of the bubble, so these two conditions are related.
If the false vacuum mass is too small, the lowest action solution
is the Hawking Moss solution,\cite{hawkmoss}
corresponding to stochastic
fluctuations over the top of the barrier rather than
tunneling through it.  This gives unacceptable density
perturbations,
 \cite{stein,linde-etal,linde-book}.

It is not clear how difficult it is to construct
potentials which can give rise
to the general thin wall
background used here.  Simple $\phi^4$ potentials in
one field models tend to result in the Hawking Moss instanton
\cite{lm} when the slow roll condition in the true vacuum is
combined with thin wall conditions.
A linear potential in both the true and false
vacuum (triangular barrier)
apparently forces the energy difference to be
negligible. In addition to finding out which parameter
values are consistent with the thin wall approximation,
it is particularly interesting
to see what potentials arise naturally in
particle physics models.

\section{Fluctuations around the background solution}
With the background solution of the previous section, eq.
(\ref{inst}),
one can now solve for basis functions in both the true and
false vacuum and match the false vacuum fluctuations across
the wall.
(A rigorous
foundation for this using the WKB formalism is found in \cite{wkb-kyot}.
See also \cite{wkb-tun,rub,kandrup,vv,TS} and references therein.)

The potential right before nucleating the bubble is assumed to
be a deep well, and after nucleating practically flat, for
the second bout of inflation.  Schematically,
\be
V(\phi) =\left\{
\begin{array}{lr}
V_F + M^2 (\phi-\phi_i)^2 + \cdots
  \; \; \; & \phi \approx \phi_i \\
-\mu^3 \phi     \; \; \; & \phi \approx \phi_n
\end{array}
\right.
\ee

For a thin wall, the matching is in region II.  The discussion of
basis functions as in \cite{BGT} is reviewed first.  The region II
fluctuations around the classical solution will be given for
Lorentzian signature and then rotated to Euclidean signature.  
They obey (since the background is constant except at the wall)
\be
(\Box -
{\delta^2 V(\phi) \over \delta \phi^2}|_{\phi = \phi_{b}})
\delta \phi = (\Box - m^2(\sigma)) \delta \phi = 0 \label{flucts}
\ee
The mass only depends upon $\sigma$ because the background
solution $\{ \phi_b(\sigma), b(\sigma) \}$
only depends upon $\sigma$.  So with a step function
at $b(\sigma) = R$, the boundary conditions are
\be
m^2(\sigma) = \left\{
\begin{array}{lr}
M^2 \; \; ,& \; R < b(\sigma) < \infty \\
0 \; \; , &\; 0 < b(\sigma) < R
\end{array}
\right.
\ee

Expanding eq. (\ref{flucts}) in terms of $(\sigma, \tau,\theta,\phi)$,
and the background de Sitter metric,
\be
[{\part^2 \over \part \sigma^2} + 3 H\cot H \sigma {\part \over \part
\sigma}
- {H^2 \over \sin^2 H\sigma}({\part^2 \over \part \tau^2}
+ 2 \tanh \tau {\part \over \part \tau}
+ {{\bf L}^2 \over \cosh^2 \tau}) - m^2(\sigma) ]
 \phi(\sigma,\tau,\theta,\phi) = 0. \label{flucteq}
\ee
The eigenfunctions of the Laplacian on the 3d hyperboloid
satisfy
\be
({\part^2 \over \part \tau^2}
+ 2 \tanh \tau {\part \over \part \tau}
+ {{\bf L}^2 \over \cosh^2 \tau})
Y_{plm}(\tau,\theta,\phi) = -(1+p^2)Y_{plm}(\tau,\theta,\phi) \; ,
\ee
so schematically
\be
\phi(\sigma,\tau,\theta,\phi) =
\int dp \, S^{\nu '}_p(\sigma)Y_{plm}(\tau,\theta,\phi) \; . \label{phisoln}
\ee
Here $Y_{plm}(\tau,\theta,\phi) = f_{pl}(\tau) Y_{lm}(\theta,\phi)$.
The $Y_{lm}(\theta,\phi)$ are the usual spherical harmonics.
Since all the mass dependence is in $S^{\nu '}_p(\sigma)$,
through
\be
\nu ' = \sqrt{{9 \over 4} - {m^2 \over H^2}} - {1\over 2} \; ,
\ee
consider $\sigma$ dependence first.

The equation of motion for $S^{\nu '}_p(\sigma)$ is
\be
[{\part^2 \over \part \sigma^2} + 3H \cot H\sigma {\part \over \part
\sigma}
+ {H^2(1+p^2) \over \sin^2 H\sigma} - {m^2(\sigma)} ]
S^{\nu '}_p(\sigma)
= 0.
\ee

The most convenient basis, different from the one usually
used to describe the tunneling as in \cite{col1}, is the
$(u, \tau)$ coordinate system used to describe the instanton in
\cite{TS}.  Recalling that
$\cos [H\sigma] = \tanh u$,
\be
S^{\nu '}_p(\cos [H\sigma]) = {F_{\nu '}^{ip}(u) \over a(u)}
\ee
where
\be
\begin{array}{ll}
F_{\nu '}^{ip}(u) =&
{1 \over 4 \pi \sqrt{|p|}} \{a_{\nu '+}
 \Gamma(1-ip)P^{+ip}_{\nu '}(\tanh u )
-{C_2\over |C_2|}a_{\nu '-}
 \Gamma(1 + i p)P^{-ip}_{\nu '}(\tanh u )\}\\
&=_{x\equiv \tanh u}
{1 \over 4 \pi \sqrt{|p|}} \{a_{\nu '+}
({1 + x \over 1-x})^{ip\over 2}
 {}_2F_1(-\nu ', \nu ' + 1; 1-ip; {1-x\over 2})
\\
&-{C_2\over |C_2|}a_{\nu '-}   ({1 + x \over 1-x})^{-ip\over
2}
{}_2F_1(-\nu ', \nu ' + 1; 1+ip; {1-x\over 2}) \} \; .
\end{array}
\label{Fsoln}
\ee
The coefficients
\be
a_{\nu ' \pm} = \sqrt{1 \pm \sqrt{1 - |C_2|^2/C_1^2} \over 2} \;
\ee
depend on momentum,
\be
\begin{array}{l}
C_1(p) = 2 \pi [1 + {\sin^2[\pi \nu '] \over \sinh^2[\pi p]}] \\
C_2(p) = 2 \pi {\Gamma(1-ip) \Gamma(ip-\nu ') \over
\Gamma(1+ip) \Gamma(-ip-\nu ')}
{\sin[\pi \nu ']\sin[\pi(\nu ' -i p)] \over \sinh^2[\pi p]} \\
= -2 \pi^2 {\Gamma(1-ip) \over \Gamma(1+ip)}
{\sin \pi \nu ' \over \sinh^2 \pi p}
{1
\over \Gamma(-ip-\nu')\Gamma(1-ip +\nu ')}
\end{array}
\ee
The coefficient $C_1(p)$ is real for values of $\nu '$ considered here,
$C_2(-p) = C_2^*(p)$ and
$\lim_{p_\rightarrow \infty , \nu ' \; fixed} C_2(p) = 0$.

The functions $F^{ip}_{\nu '}$ are normalized as
\be
\int_{-\infty}^{\infty} du F_{\nu '}^{ip}(u)
F_{\nu '}^{i p' *}(u) = {1 \over 8 \pi
|p|}
\delta(p-p') \; .
\label{fnorm}
\ee

There is a continuum of modes for $0 \le p \le \infty$.
(The $\tau$ dependence introduces some more $p$ dependence
so that the full wavefunctions are nonsingular at $p=0$.)

For the special cases of $\nu ' =0$ (conformal mass, $M^2 = 2H^2$)
and $\nu ' = 1$ ($M^2 = 0$ as is the case after tunneling), one has
\be
\begin{array}{l}
F_0^{ip} = {e^{i p u} \over 4 \pi \sqrt{|p|}} \\
F_1^{ip} = {e^{ipu}\over 4 \pi \sqrt{|p|}}{(\tanh u - ip)
\over 1-ip}
\end{array}
\label{exampleF}
\ee

When $0 \le \nu ' \le 1$ (ie when $M^2/2H^2 <1$),
there is an additional normalizable
state which was shown to be part of the basis of
states in \cite{vacmode}.
They showed that it is necessary to include this state in
order to obtain the
proper Wightman function for the
Bunch-Davies vacuum of de Sitter space.

The Wightman function is the sum over positive frequency modes
times the negative frequency modes (a vacuum dependent
assignment),
\be
\begin{array}{ll}
G^+(x,x') &= \langle 0|\hat{\phi}(x) \hat{\phi}(x')|0\rangle  \\
&=\langle 0|\sum_{k,k'} \{(u_{+,k}(x)\hat{a}(k) + u_{-,k}(x)
 \hat{a}^\dagger(k))\}\\
& \; \; \; \; \; \; \; \times
 \{ (u_{+,k'}(x')\hat{a}(k') + u_{-,k'}(x')
 \hat{a}^\dagger(k')) \}
|0\rangle \\
&= \sum_k u_{+,k}(x)u_{-,k}(x') \; .
\end{array}
\ee
The modes $u_{\pm,k}(x)$ obey the Klein Gordon equation and
are normalized according to
\be
\begin{array}{ll}
(u_{+,k}(x), u_{+,k'}(x)) &= \delta (k-k') \\
(u_{-,k}(x), u_{-,k'}(x))&=- \delta (k-k') \\
(u_{+,k}(x), u_{-,k'}(x)) &= 0 \; .
\end{array}
\ee
where the inner product is the Klein Gordon inner product:
\be
(\phi_1(x), \phi_2(x)) =
-i \int_{\Sigma} d\Sigma^\mu  [-g_{\Sigma}(x)]^{1\over 2}\{\phi_1(x)
\part_\mu \phi_2^*(x)
-\part_\mu \phi_1(x) \phi_2^*(x) \} \; \; . \\ \label{kg}
\ee
The integration is over $\Sigma$, a
 (globally) Cauchy surface,
{\it i.e.}, a spacelike hypersurface
which every non-spacelike curve intersects exactly
once\cite{HawkE}.

The additional normalizable state for $0 < \nu ' \le 1$ has
$p=i \nu '$,
\be
F_{\nu '}^{-\nu '} = {1 \over 4 \pi \sqrt{|p|}}
(\cosh u)^{- \nu '} \; \; \; 0 < \nu ' \le 1 \;   \label{discnorm}
\ee
It is for a discrete value of $p$ and hence corresponds to a
state which normalizes to a constant times a Kronecker
delta function, not a Dirac $\delta$ function.

There is a discussion of the properties of these states in
\cite{vacmode}.  When
continued to the inside of the light cone of the bubble center, region I,
these states
appear as supercurvature modes which are naively
unnormalizable.  As explained in \cite{vacmode},
this lack of normalizability
is due to the curves of
fixed time $T$
failing to be Cauchy surfaces.  (The light
cone of the bubble center, a non-spacelike curve,
is $T=0$, and never intersects curves with $T\ne 0$.)
They show that for the other modes, with $p^2 >0$,
the inner products on curves of
fixed time $T$ coincide with their norms on Cauchy surfaces
because of sufficiently fast falloff at infinity.
The supercurvature modes
are normalizable on a Cauchy surface, for example along the
$\tau = 0$ surface outside the forward light cone\cite{vacmode},
and the result is independent of Cauchy surface as it should be.

The $\ell = 0$ mode for $\nu ' \rightarrow 1$ has zero norm, as shown
in \cite{vacmode},
perhaps due to overlap with a translational zero mode
in the system, similar to what happens in \cite{gar96}.
Aside from this case, the normalizable
discrete state with special $p = i \nu '$
occurs as part of the
Bunch-Davies vacuum for all $\ell,m$, when $0<\nu ' \le 1$.

When $\tau$ dependence is included,
the Wightman function can be used, via the Klein Gordon
inner product, to project upon positive or negative frequency
modes for a given set of basis functions.  This was used
in \cite{BGT,BT} to identify the $\ell = 0$
positive freqency mode functions
for the Bunch Davies vacuum
\be
\phi^+_{\pm p}(u,\tau) =
 H \cosh u F_{\nu '}^{\pm ip}(u){(e^{\pi p\over 2}e^{-ip\tau} -
 e^{-\pi p\over 2} e^{ip\tau})
\over\sqrt{e^{\pi p} - e^{-\pi p}} \cosh \tau} \;
 \; 0 < p < \infty
\label{posf}
\ee
with corresponding negative frequency modes
\be
\phi^-_{\pm p}(u,\tau) =
H \cosh u F_{\nu '}^{\pm ip}(u){(e^{-\pi p \over 2}e^{-i p\tau} -
e^{\pi p\over 2} e^{ip\tau})
\over \sqrt{e^{\pi p}-e^{-\pi p}}\cosh \tau} \; \; 0 < p < \infty \; .
\ee
For general $\ell$, the
$\tau$ dependent
factor $f_{pl}(\tau )$ is proportional to (from \cite{vacmode}):
\be
\begin{array}{ll}
f_{pl}(\tau) & \propto {\Gamma(ip +l +1) \over \Gamma(ip)}{1 \over \sqrt{\sinh
(\tau + \pi i / 2)}} P^{-l - 1/2}_{ip-1/2}(\cosh (\tau + \pi i /2)) \\
  &= i(-1)^{l+1} \sqrt{2 \over \pi} p{\Gamma(-ip) \over \Gamma(-ip +l +1)}
\cosh^l\tau \left({d \over d \sinh \tau}\right)^l{\sin p (\tau + i \pi/2)
\over \cosh \tau} \; \; .\\
\end{array}
\ee
For simplicity, the discussion will be restricted hereon to $\ell=m =
0$,
since the matching across the bubble wall
 depends on $\sigma$ and not $\tau$.
So, suppressing $\ell, m$ dependence,
the operator describing fluctuations is
\be
\begin{array}{ll}
\delta \hat{\phi}(u,\tau) =&
\int_0^\infty dp \{ \phi^+_{ p}(u,\tau) \hat{a}(p)
+ \phi^+_{- p}(u,\tau) \hat{a}(-p) + h.c. \} \\
& + \sum_{p_{disc}} \phi^+_{ p_{disc}}(u,\tau) \hat{a}(p_{disc})
  + h.c.  \label{phiexp}
\end{array}
\ee
with the $\phi^{\pm}_{\pm p}$ normalized in the
Klein-Gordon inner product, eqn (\ref{kg}).

\section{Initial conditions, matching across wall and normalization}
Given the positive frequency modes in the false vacuum, the
next step is to analytically continue to Euclidean time and
match them on to the initial conditions for fluctuations
around the instanton,
eq. (\ref{inst}).
However, the initial conditions for the
instanton are not directly
visible in the
$(\sigma,\tau_E)$ or $(u,\tau_E)$ coordinate system.
In these coordinates,
the bubble
wall is present for all time $\tau_E$,
so there is no false vacuum at any time $\tau_E$.  However,
these coordinates are useful since
a constant value of $b(\sigma),a(u)$ defines the bubble wall.

This problem is emphasized in \cite{TS}, where another
coordinate system is suggested which maps
$(\sigma,\tau_E) \rightarrow(\tilde{\sigma},\tilde{\tau}_E)$.
The new time coordinate $\tilde{\tau}_E$ coincides with $\tau_E$
along the line
$\tau_E =0$, the nucleation time,
where analytic continuation to Lorentzian
signature takes place.
The $(\tilde{\sigma},\tilde{\tau}_E)$
coordinate system can be used to specify
initial conditions, but it is not useful in practice for matching across
the bubble wall.
At early
times $\tilde{\tau}_E$ in the $(\tilde{\sigma}, \tilde{\tau}_E)$
coordinate system,
the bubble wall has not yet formed, so that
the vacuum can be identified.
One can do the matching
using either coordinate
system.  The modes corresponding to the false vacuum
Bunch-Davies modes in the $(\sigma, \tau_E)$ coordinates
coincide with those in the far past, at early times
$\tilde{\tau}_E$, in the coordinate system which has
the proper initial conditions (false Bunch-Davies vacuum).
So, knowing that the $(\tilde{\sigma},\tilde{\tau})$ coordinate
system exists justifies starting with false vacuum Bunch-Davies
modes exterior to the bubble, as these are continued from modes
at early
$\tilde{\tau}_E$.\footnote{I thank Arley Anderson for discussions about
this.}

An important consequence of
not specifying the initial conditions
at fixed $\tau_E$ (the bubble wall exists for
all $\tau_E$, the initial condition is without the bubble),
that is, using the basis functions eq. (\ref{posf}) for matching,
is that the resulting wavefunction matched across
the bubble wall is not automatically
normalized.  So the wavefunction is normalized after the
nucleation time $\tau_E = 0$.  The Minkowski space
analogue of this was done
in, {\it e.g.}, \cite{yts}.

With this understood, the first step is to analytically
continue the positive frequency wavefunctions (\ref{posf})
of the false vacuum to Euclidean time,
$\tau \rightarrow i \tau_E$,
\be
A_{p,\nu '}(u,\tau_E) =
\phi^+_{\pm p}(u,\tau_E) = H \cosh u F_{\nu '}^{\pm ip}(u)
{(e^{\pi p \over 2} e^{p\tau_E} - e^{-\pi p\over 2} e^{-p\tau_E})
\over\sqrt{e^{\pi p} - e^{-\pi p}} \cos \tau_E} \; \; .
\ee
Recall
$\nu ' = \sqrt{{9 \over 4} - M^2}-{1 \over 2}$,
and $F_{\nu '}^{ip}(u)$ is given in eq. (\ref{Fsoln}).
These are
to be matched at $u = u_R$ onto linear combinations of
modes corresponding to the true vacuum, where
$M^2 \approx 0$, that is $\nu ' \sim 1$.
The basis functions for true vacuum modes are
\be
B_{\pm p}(u, \tau_E) =  H_T \cosh (u+\delta)
F_1^{\pm ip}(u+\delta)
{(e^{ \pi p \over 2}e^{ p\tau_E}-
 e^{- \pi p\over 2} e^{- p\tau_E})
\over 2 \sqrt{e^{\pi |p|} - e^{-\pi|p|}} \cos \tau_E} \; \; .
\ee
where $F^{ip}_1(u)$ is given in eq. (\ref{exampleF}).

Because the $\tau_E$ dependence is continuous,
one can only match modes with equal values of $|p|$ (the
matching has to be true for all times, so the time dependence has
to be the same on both sides).  Both
$B_{\pm p} (u, \tau_E)$ are positive frequency modes, so
this does not look like particle
creation.
Continued inside the light cone,
where the space and time
dependence get interchanged, it can be interpreted as
causing particle creation.
Matching the functions and
their first derivatives at $u = u_R$,
\be
\begin{array}{l}
A_{p}(u, \tau_E) = \alpha_p B_{ p}(u, \tau_E) +
\beta_p B_{-p}(u , \tau_E) \\
\partial_u A_{p}(u, \tau_E) = \alpha_p \partial_u B_{ p}(u, \tau_E) +
\beta_p \partial_u B_{-p}(u, \tau_E) \; ,\\
\end{array}
\ee
 gives
\be
\begin{array}{l}
\alpha_p = {\partial_u B_{-p} A_{p} - B_{-p} \partial_u A_{p} \over
B_p \partial_u B_{-p} - B_{-p} \partial_u B_{p}} \\
\beta_p = {-\partial_u B_{p} A_{p} + B_{p} \partial_u A_{p}
\over B_p \partial_u B_{-p} - B_{-p} \partial_u B_{p}} \\
\end{array} \label{bogcoeff}
\ee
evaluated at $u = u_R$.

So for false vacuum mass $M^2 = 2H^2$, (using the explicit form
of $F^{ip}_0(u)$ in eq. (\ref{exampleF})),
\be
\begin{array}{ll}
\alpha_p &= {e^{-ip \delta} \over 2 p(p-i)}
[ 2 p^2 + {\cosh \delta - i p \sinh (2u_R+\delta)
\over \cosh u_R \cosh(u_R+\delta)} ] \\
\beta_p& = {e^{ip (2u_R+\delta)} \over 2 p(p+i) }
{\cosh \delta -ip \sinh \delta \over
\cosh u_R \cosh (u_R+\delta) } \\
\end{array} \label{bogcoef}
\ee
For more general masses in the false vacuum,
the Bogoliubov coefficients are:
\be
\begin{array}{ll}
\alpha_{p,\nu '} &={ e^{-ipu_R} \over 2 p(p-i)}
 \{
 (1 + p^2 -(\nu' +2)\tanh u_R(\tanh (u_R+\delta)+ip))
4 \pi\sqrt{|p|}F_{\nu '}^{ip}(u_R)\\
&+ (\nu ' -ip +1)(\tanh (u_R+\delta) +ip)
 a_{\nu '+} \Gamma(1-ip)P^{ip}_{\nu ' +1}(\tanh u_R) \\
&-(\nu ' + ip +1)(\tanh (u_R+\delta) + ip)
 {C_2\over |C_2|}a_{\nu '-}
\Gamma(1 + i p)P^{-ip}_{\nu '+1}(\tanh u_R) \} \\
\end{array} \label{alphabog}
\ee
and
\be
\begin{array}{ll}
\beta_{p,\nu '} &=
{ e^{ip(u_R+\delta)} \over 2p(p+i)} \{
( 1+ p^2 - (\nu ' +2)\tanh u_R ( \tanh (u_R+ \delta)-ip ))
 4 \pi \sqrt{|p|}F_{\nu '}^{ip}(u_R)\\
&+
(\nu ' -ip +1)(\tanh (u_R+\delta) -ip) a_{\nu '+}
 \Gamma(1-ip)P^{ip}_{\nu ' +1}(\tanh u_R) \\
&-
(\nu ' + ip +1)(\tanh (u_R+\delta) - ip) {C_2\over |C_2|}a_{\nu '-}
\Gamma(1 + i p)P^{-ip}_{\nu '+1}(\tanh u_R) \} \; .
\end{array} \label{betabog}
\ee
One can verify that as the radius goes to zero,
$u \rightarrow \infty, \delta \rightarrow 0$,
and these reduce to the expressions of \cite{BGT,BT},
\eq
\alpha_p \rightarrow_{R\rightarrow 0} a_{\nu '+} \; ,
\; \; \; \;
\beta_p \rightarrow_{R\rightarrow 0} -{C_2\over |C_2|} a_{\nu ' -}
\en
In showing this it is helpful to recall that
$\lim_{u\rightarrow \infty} \Gamma(1-ip) P^{ip}_{\nu '}(\tanh u)
\sim e^{ipu}$,
so that terms cancel between $F^{ip}_{\nu '}$ and $P^{ip}_{\nu ' +1}$.
In addition, as $p \rightarrow \infty$ these go over to the
conformally coupled $M^2 = 2 H^2$ case,
\be
\alpha_{p,\nu '} \rightarrow 1 \; , \beta_{p,\nu '} \rightarrow 0 \; .
\ee
That is, at short distances, the dependence
upon the scales set by the mass and radius drops out.

These Bogoliubov coefficients are for $p$ real.  When $0 < \nu ' \le 1$,
as mentioned earlier, there are
also states corresponding to supercurvature modes in the
false vacuum. These
supercurvature modes have $\tau$ dependence which
match onto eigenfunctions of the Laplacian with $u$ dependence
\be
\begin{array}{l}
F^{\nu '}_1 (u+\delta) = e^{-\nu ' (u+\delta)}
{(\tanh (u+\delta) + \nu ') \over 1 + \nu '}  \\
or \; F^{-\nu '}_1 (u+\delta) = e^{\nu ' (u+\delta)}
{(\tanh( u+\delta) - \nu ') \over 1 - \nu
'}  \; \; ,
\end{array}
\label{supmatch}
\ee
non-normalizable on any surface.  The
first
eigenfunction is well behaved as $u\rightarrow \infty$ and can be
connected across the wall at $u = u_R$ to eigenfunctions
which are well behaved at $u=- \infty$, however,
it is not possible generically
to match the original vacuum
mode and its first derivative (two constraints)
across the wall without the second eigenfunction which
blows up at $u_R = \infty$.
The power spectrum inside the bubble may be computed
anyhow, {\it e.g.} as is shown in
\cite{lyth} for general supercurvature modes,
but it is possible that the lack of normalizability means
that there is some instability (and thus inconsistency in
this naive treatment).  The way out of this is \cite{new}
to not include the state which has become
unnormalizable due to the wall, but to look for any
new states that may have become normalizable at
the same time, again due to matching across the wall.
Thus one has an added degree of freedom (the value
of $p$) which can be used to try to satisfy the
matching constraints.
As they show, a normalizable supercurvature
state is present for $M^2/H^2 \le 1$ for
$R=0$, and for larger values of $M$ when $R \ne 0$. 

Similarly, there is also a
supercurvature mode present in the 'true' vacuum of the bubble
interior, where $\nu ' \approx 1$.  It does not have the same time
dependence
as any state in the false vacuum, and so is not excited by
the continuation of the false vacuum fluctuations across the wall.
It is present in some cases, {\it e.g.}
if there is no bubble, \cite{vacmode,BY}, or if the field
providing density fluctuations is massless on both sides of the
bubble wall \cite{cmb}.
If the mass of the field in the false vacuum is large enough,
the question of
these supercurvature modes
does not occur.
These modes will not be considered further here.

Once the modes have been matched across the wall, the
wavefunctions are analytically continued to Lorentzian signature
at $\tau_E = 0 = - i \tau$
to give the wavefunctions for a bubble outside
of the light cone of the center of the bubble.
This is straightforward to do by substituting $\tau = i \tau_E$.

The resulting wavefunctions are
\eq
\phi^+_{+p,{\rm unnorm}}(u,\tau) =
\theta(u_R-u)A_{p,\nu '}(u,\tau) +
\theta(u-u_R)( \alpha_p B_{ p}(u, \tau) +
\beta_p B_{-p}(u , \tau)) \; ,
\label{ unnorm}
\en
where $\theta(u)$ is a step function.
As mentioned before,
 $\phi^+_{+p,{\rm unnorm}}(u,\tau)$ unnormalized.  In fact, it is
not even\footnote{I thank
M. Bucher for pointing this out and for suggesting the method
used below.} orthogonal
to $\phi^+_{-p,{\rm unnorm}}(u,\tau)$.
This can be remedied by using the same method in \cite{BT}.

As the $\tau$ dependence is the same for both fields, and
corresponds to a properly normalized wavefunction
(as would be automatic if there were no bubble wall, which
also has the same $\tau$ dependence), it can be separated out
first,
\eq
 \phi^+_{+p,{\rm unnorm}}(u,\tau) = {{\cal F}^{+p}(u) \over a(u)}
{(e^{\pi p \over 2} e^{p\tau_E} -
e^{-\pi p\over 2} e^{-p\tau_E})
\over 2 \sqrt{e^{\pi |p|} - e^{-\pi|p|}}\cos \tau_E} \; \; .
\en

Then, the ${\cal F}^{\pm p}(u)$
are eigenfunctions of a self-adjoint operator with eigenvalue
depending on $p^2$, and so eigenfunctions with different eigenvalues
are orthogonal,
\eq
\int_{-\infty}^\infty \, du {\cal F}^{+p}(u){\cal F}^{+p'}(u)
={D_1(p) \over 8 \pi |p|} \delta(p+p') + {D_2(p)\over
8 \pi |p|} \delta (p-p')
\en
The coefficients
$D_1(p), D_2(p)$ can be read off from the asymptotic behavior at
$u \rightarrow \pm \infty$.   After some algebra they are
seen to be:
\eq
\begin{array}{ll}
D_1(p) &= {1 \over 2} \left[ |\alpha_p|^2 + |\beta_p|^2 +
1 \right] = D_1(-p) \\
D_2(p) &= \left[ \alpha_p \beta_{p}
+
{C_2\over 2 C_1}  \right]= D_2(-p)^*\\
\end{array}
\en

Then the normalized wavefunction is
\eq
\phi^+_{+p}(u,\tau) =
b_{+,p}(u_R) \phi^+_{+p,{\rm unnorm}}(u,\tau)
-{D_2(p) \over |D_2|} b_{-,p}(u_R)\phi^+_{-p,{\rm unnorm}}(u,\tau)
\label{phinp}
\en
where
\eq
b_{\pm,p}(u_R) =
\sqrt{D_1 \over D_1^2 - |D_2|^2}
 \sqrt{1 \pm \sqrt{1 - |D_2|^2/D_1^2} \over 2} \; \; .
\en

This normalization can be checked by a different calculation once
the modes are continued into the future light cone of the
bubble center, as described below.
The other method of normalization has no integration
over (the analytic continuation of) $u$, and so does
not depend on continuity the same way.  The two calculations agree.
So the operator describing fluctuations, $\delta \hat{\phi}$, is
as in eq. (\ref{phiexp}), with $\phi^+_{p}(u,\tau)$
given in eq. (\ref{phinp}).

\section{Continuation into open universe, power spectrum}
The operator for fluctuations is next continued inside
region I to find its effect in the open universe.
For $u>u_R$, approaching the light cone,
\eq
\begin{array}{ll}
\phi^+_{p}(u,\tau)|_{u>u_R} &=
(b_{+,p}(u_R) \alpha_p -{D_2(p)\over|D_2|}
 b_{-,p}(u_R) \beta_{-p})B_p(u,\tau) \\
&+ (b_{+,p}(u_R) \beta_p -{D_2(p)\over|D_2|}
 b_{-,p}(u_R) \alpha_{-p})B_{-p}(u,\tau)\\
\end{array}
\en
At the light cone, $\sigma \rightarrow 0$ ($u \rightarrow \infty$).
Inside region I,
$\sigma = - i T, \tau = \chi - i \pi/2$, and time and space
get interchanged.  The analogue of $u$ in region II is the
conformal coordinate $\eta$, obeying $e^{\eta} =\tanh(T/2)$.

The $\sigma,\tau$ coordinate system becomes singular at the
light cone.
The correspondence of wavefunctions across the light cone was found in
\cite{BGT}, and can be
taken over directly, as they
emphasize,
since only the short distance behavior of the fields is used:
\be
\begin{array}{l}
e^{-i p u} \cosh u {e^{i p \tau} \over \cosh \tau} \rightarrow
(+i) {\sin p \chi \over \sinh \chi} e^{(ip - 1)\eta} \\
e^{-i p u} \cosh u {e^{-i p \tau} \over \cosh \tau} \rightarrow
0 \\
e^{i p u} \cosh u {e^{i p \tau} \over \cosh \tau} \rightarrow
0 \\
e^{i p u} \cosh u {e^{-i p \tau} \over \cosh \tau} \rightarrow
(-i) {\sin p \chi \over \sinh \chi} e^{-(ip + 1)\eta} \\
\end{array}
\label{across}
\ee
Only half of the modes propagate inside the light cone for massless fields,
the other propagate along the light cone instead \cite{BGT}.
Since the matching depends only on the coefficients of $i \tau$ in
the exponential, these rules can be used directly to match the modes for
general $\ell$ if desired.

The basis functions then become:
(one can think of $u$ here as $u + \delta$ in the
earlier basis functions)
\be
\begin{array}{ll}
B_{p}(u ,\tau)
& \rightarrow_{u \rightarrow \infty}
-iH_T{1\over 4 \pi \sqrt{p}}
{e^{\pi p \over 2}  \over
\sqrt{e^{\pi p}-e^{-\pi p}}}
e^{(-ip-1)\eta}
{\sin p\chi \over \sinh \chi}
\equiv \tilde{B}_{p}(\eta,\chi) \\
B_{-p}(u ,\tau)
&\rightarrow_{u \rightarrow \infty}
-iH_T{1\over  4 \pi
\sqrt{p}}
{ e^{-\pi p \over 2}  \over
\sqrt{e^{\pi p}-e^{-\pi p}}}
e^{(ip-1)\eta}{\sin p\chi \over \sinh \chi}
\equiv \tilde{B}_{-p}(\eta,\chi) \; .
\end{array}
\label{bacross}
\ee

With this continuation across the future light cone
of the bubble center, eq. (\ref{phiexp})
becomes
\be
\begin{array}{ll}
\delta \hat{\phi}(\eta,\chi)
= &\int_0^{\infty}
dp \,
\left[(b_{+,p}(u_R) \alpha_p -{D_2(p)\over|D_2|}
 b_{-,p}(u_R) \beta_{-p})\tilde{B}_p(\eta,\chi)
\right. \\
&\left. + (b_{+,p}(u_R) \beta_p -{D_2(p)\over|D_2|}
 b_{-,p}(u_R) \alpha_{-p})\tilde{B}_{-p}(\eta,\chi)
\right] \hat{a}^+(+p)\\
&+\left[(b_{+,p}(u_R) \alpha_{-p} -{D_2(-p)\over|D_2|}
 b_{-,p}(u_R) \beta_{p})\tilde{B}_{-p}(\eta,\chi)\right. \\
&
\left. + (b_{+,p}(u_R) \beta_{-p} -{D_2(-p)\over|D_2|}
 b_{-,p}(u_R) \alpha_{p})\tilde{B}_{p}(\eta,\chi)
\right] \hat{a}^+(-p)\\
&+ h.c.
\end{array}
\label{phitrue}
\ee
(Recall there are no discrete states for the values of $M^2$
being considered.)

This analytically continued wavefunction can be verified to have
proper normalization, providing a check on the calculation
of normalization given earlier.  The future in de Sitter space
has two light cones which must be included, region I, where
the open universe interior to the bubble is found, and that of
the antipode of the center of the bubble (for descriptions
see, {\it e.g.} \cite{BGT,vacmode,ballen}), region III.
(These are regions $L,R$ in the notation of \cite{vacmode}.)
For our purposes,
continuing into region III can be done by using symmetry
and then verifying that a normalized wavefunction in
region II remains normalized when continued into region $I,III$
using these rules.
The modes continued into region III are exterior
to the bubble wall, and consequently have
 mass corresponding to the false vacuum
mass.
As time and space are
interchanged in these forward light cones,
the Klein Gordon inner product now involves Wronskians of the
analytically continued $u$ dependent part of the
wavefunctions.
The inner product is calculated on a fixed time
surface in both forward light
cones and then added.  Although, as stated before, the fixed time
surfaces are not Cauchy surfaces, the wavefunctions under
consideration are the same ones (restricted to each light cone)
discussed in \cite{vacmode}.  They
do not correspond to supercurvature modes and consequently
have sufficiently
fast falloff at infinity that their inner products
on the fixed
time
surface inside the light cones
correspond to their inner products on a Cauchy surface.
It is also useful to know, in order to show this, that
$
|\alpha_p|^2 - |\beta_p|^2 = a_{\nu ' +}^2 - a_{\nu ' -}^2 \;
$
which can be shown with the definitions in eq. (\ref{bogcoeff})
and
properties of Wronskians.

The fluctuation $\delta\hat{\phi}$
inside the light cone sources gravity, and
can be matched
directly onto fluctuations of the
gravitational
potential $\Phi$ using results of \cite{BGT},
as follows.
Since the fluctuations are small, they can be described
by expanding the metric around de Sitter space and
the field $\phi$ around its classical value, as was done
in \cite{BGT}, also see \cite{BY} for another discussion.
The result is that a source $\phi \approx e^{(\pm ip -1) \eta}$
corresponds to a ``gauge invariant gravitational potential''
\be
\Phi \approx {4 \pi G V_{,\phi} \over \pm i p +2}e^{(\pm i p +1) \eta}
\; .
\ee
Then, as \cite{BGT} show, when $H$ remains constant
and the potential is linear, this matches onto the exact solution:
\be
\Phi = {4 \pi G V_{,\phi} \over \pm i p +2}e^{(\pm i p +1) \eta}
[1 - {p \pm i \over 3(p \mp i)}e^{2 \eta}] \; .
\ee

For the $\delta\hat{\phi}$ found here, the corresponding positive
frequency $\hat{\Phi}$ is:
\be
\begin{array}{ll}
\hat{\Phi}(\eta,\chi) &=-i {4 \pi G V_{,\phi} \over 4 \pi H_T}e^\eta
[\int_0^\infty {dp \over \sqrt{p}} {\sin p \chi \over \sinh \chi}
{1 \over \sqrt{2 \sinh  p \pi}} \\
&\{ (e^{\pi p \over 2}
(b_{+,p}(u_R) \alpha_p -{D_2(p)\over|D_2|}
 b_{-,p}(u_R) \beta_{-p})
{1 \over 2 - ip}
e^{-ip\eta} [1 -{p-i \over 3(p+i)}e^{2\eta}] \\
&+
e^{-\pi p \over 2}
(b_{+,p}(u_R) \beta_p -{D_2(p)\over|D_2|}
 b_{-,p}(u_R) \alpha_{-p})
 {1 \over 2+ ip}e^{ip\eta}
[1 -{p+i \over 3(p-i)}e^{- \eta}] ) \hat{a}(p)\\
&+(e^{-\pi p \over 2}
(b_{+,p}(u_R) \alpha_{-p} -{D_2(-p)\over|D_2|}
 b_{-,p}(u_R) \beta_{p})
{1 \over
2-ip}e^{ip\eta}[1-{p+i \over 3(p-i)}e^{2\eta}] \\
&+ e^{\pi p \over 2}
(b_{+,p}(u_R) \beta_{-p} -{D_2(-p)\over|D_2|}
 b_{-,p}(u_R) \alpha_{p})
 {1\over 2-ip}
e^{-ip\eta}[1-{p-i \over 3(p+i)}e^{2 \eta}])\hat{a}(-p)
 \}
\end{array}
\ee

The power is defined by taking
\be
\langle \Phi(\chi = 0, \eta) \Phi(\chi,\eta ) \rangle
= \int_0^\infty dp \, p{\sin p\chi \over \sinh \chi}P_{\Phi}(p,\eta)
\; + discrete \; states
\ee
and isolating the growing mode by
taking $\eta \rightarrow 0^-(T\rightarrow \infty)$.
This gives
\be
\begin{array}{ll}
P_{\Phi}(p) =& (G V_{,\phi})^2{1 \over p(1+p^2) 2\sinh \pi p}{4 \over
9 H_T^2}
\\
&\{ |e^{\pi p \over 2}
(b_{+,p}(u_R) \alpha_p -{D_2(p)\over|D_2|}
 b_{-,p}(u_R) \beta_{-p})
\\
&
-{p+i \over p-i}
e^{-\pi p \over 2}
(b_{+,p}(u_R) \beta_p -{D_2(p)\over|D_2|}
 b_{-,p}(u_R) \alpha_{-p}) |^2 \\
&+|(e^{-\pi p \over 2}
(b_{+,p}(u_R) \alpha_{-p} -{D_2(-p)\over|D_2|}
 b_{-,p}(u_R) \beta_{p})
\\
&-
{p-i\over p + i}
e^{\pi p \over 2}
(b_{+,p}(u_R) \beta_{-p} -{D_2(-p)\over|D_2|}
 b_{-,p}(u_R) \alpha_{p})|^2 \\
&= (G V_{,\phi})^2{1 \over p(1+p^2) 2\sinh \pi p}{4 \over
9 H_T^2} [e^{\pi p} + e^{-\pi p} \\
&-{1\over D_1^2 - |D_2|^2} \{
{p+i \over p-i} (\alpha_{-p}
\beta_p - \beta_p^2 {C_2^* \over 2 C_1} -
\alpha_{-p}^2 {C_2 \over 2 C_1})
+
{p-i \over p+i} (\alpha_{p}
\beta_{-p} - \beta_{-p}^2 {C_2 \over 2 C_1} -
\alpha_{p}^2 {C_2^* \over 2 C_1})
 \}]
\label{powergen}
\end{array}
\ee
Here, the identity
\eq
\begin{array}{l}
D_1^2 - |D_2|^2 =  D_1 -  {C_2^* D_2 + C_2 D_2^* \over
2 C_1} \\
= {1\over 2}
(1 + |\alpha_p|^2 + |\beta_p|^2 -{C_2^* \over C_1}{\alpha_p \beta_p}
-{C_2 \over C_1}\alpha_{-p} \beta_{-p} - |{C_2 \over C_1}|^2)
\end{array}
\en
was used.

Taking $R \rightarrow 0$, {\it i.e.}
$\alpha_p \rightarrow a_{\nu ' +}, \;
\beta_p \rightarrow -{C_2\over |C_2|} a_{\nu ' -}, $ this
reduces to the case found in \cite{BT}:
\eq
(G V_\phi)^2 {4 \over 9 H^2} {1 \over p(p^2+1)2 \sinh \pi p}
\left[e^{\pi p} + e^{-\pi p} +
\left({C_2\over C_1}{p+i\over
p-i} +{C_2^* \over C_1}{p-i\over p+i}\right)\right] \; .
\en

Generally $|{C_2 \over C_1}| \le 1$, so in the $R=0$ case,
this means that
the $p$ dependence is in an envelope between
$\coth{ \pi p\over 2}$ and $\tanh {\pi p \over 2}$, going to
$(p(p^2+1))^{-1}$, the scale invariant
spectrum, approximately when $p\ge 2$.

For $0 \le p \le 2$ a wide range of values are possible, depending
upon the specific values of $\delta, R, M/H$.  For $p \ge 2$,
when the radius is nonzero, the last term (the last line) can only
be significant if it grows faster than the prefactor
$\sinh \pi p^{-1}$ goes to zero, in such a way as to be
comparable to the first term
 $e^{ \pi p}> 250 $.

The magnitude of this term can be bounded as follows.
It can be rewritten as
\eq
4 Re[ e^{i \phi}{p+i\over p-i}
{y A - {\epsilon_c \over 2}[1 +y^2 A^2]
\over
(\sqrt{1-\epsilon_c^2}(y^2 - 1) + y^2 + 1)A
-y\epsilon_c(A^2 + 1)
} ]
\label{lasterm}
\en
where
\eq
\begin{array}{l}
\epsilon_c = |{C_2 \over C_1}| \\
A = {C_2 \over |C_2|}{\beta_{-p} \over |\beta_{-p}|}{\alpha_{-p}\over
|\alpha_{-p}|} \\
y={|\alpha_{p}| \over |\beta_{p}|} \\
\end{array}
\en
and it has been used that $|\alpha_p|^2 - |\beta_p|^2 =
\sqrt{1-\epsilon_c^2}$ and
that ${C_2 \over C_1} = {C_2 \over |C_2|} \epsilon_c$.
(Also
recall that $\alpha_{-p} = \alpha_p^* \; , \beta_{-p} = \beta_p^*$.)
The phase $\phi$ is real, $e^{i \phi} = 
{\beta_{p} \over |\beta_{p}|}{\alpha_{-p} \over |\alpha_{-p}|}$,
and will not be of interest here since the magnitude will be
shown to be bounded and small.
{}From their definitions above, one has
$0 \le \epsilon \le 1, \; 1 \le y $ and $|A| = 1$.

It is possible to extremize the above expression in $A$ and $y$
because the extrema of a ratio of two quadratic polynomials
is given by a quadratic equation and can be solved analytically.
However, this is of
limited use.
In the space spanned by physical
values of $\epsilon_c,y,A$, the expression was not found to
ever be larger than two in absolute value (fixing one parameter,
plotting for the full range of the other two, and changing the
parameter that was originially fixed).
In the limiting value of $\epsilon_c = 1$,
the magnitude can be seen directly to be two and
when $\epsilon_c \rightarrow 0$ it is of magnitude less than two.
For any mass value, as $p$ is changed, one runs through a set of
allowed values of $\epsilon_c,y,A$, but for $p \ge 2$ this term
is effectively irrelevant, and the power spectrum becomes
 scale
invariant to a high degree of accuracy.  That is, if equation
\ref{lasterm} is less than or equal to two in magnitude, 
then the quantity in brackets
in equation \ref{powergen} is between $\coth \pi p/2$ and
$\tanh \pi p/2$, just as in the zero radius case.

The effects from the vacuum fluctuations propagating through the
bubble wall must be combined with other fluctuations caused by
the presence of the wall.
In quantizing the field theory around the classical bubble solution
$\phi_b(\sigma )$
in flat space,
there are two special modes
arising, which remain even in the thin
wall limit \cite{gv}.
One is a translational zero mode, $\part_\mu \phi_{b}(\sigma)$,
where $\part_\mu=
\part_{x_i}$, and
and one is a mode coming from
shifting the radius, $\part_\sigma \phi_{b}(\sigma)$.
The latter state was identified by \cite{HSTY} as having $p^2 = -4$
in the basis functions
of eq. (\ref{phisoln}),
and appears also in the
 de Sitter case if one assumes that
$H$ changes negligibly during tunneling \cite{gar96}.
The former
corresponds\cite{HSTY} to
(nonorthogonalized) $\ell=0,1$ modes for this value
of $p$.
As the explicit form of both
is related to derivatives of the classical
background, which only varies significantly on or near the wall,
they tend to be localized on
the wall, and
their degrees of
freedom can be related to fluctuations of the wall.
Although localized,
fluctuations in the wall can, at later times, change the definition of
where the `time' is zero.  So these fluctuations can
produce an asymmetry in space
now, since different regions would be at slightly different times in
their evolutions.
In \cite{lm,linde,juan,gar96} the bubble wall fluctuations are
discussed and
contributions to asymmetries and
to temperature anisotropies in the CMB calculated.
\section{Conclusions}
In this paper the results of \cite{BGT,BT} were extended to include
a finite wall radius and small vacuum energy difference before and
after tunneling.
The power spectrum appears to generally go over at $p \sim 2$
to an approximately scale invariant spectrum just as in the
zero radius case.
As most of the discriminating
effects found here are at very large scales,
cosmic variance may limit their
observability.   The scale invariant
spectrum that seems to arise at smaller 
scales (no exceptions were found and perhaps it
can be shown rigorously\footnote{The corresponding expression
in \cite{new} seems to be within the bounds conjectured
here.})
means that
varying the
parameters $R,\delta$ and false vacuum mass $M$ over a large 
range 
for thin wall bubble models gives a generic prediction.

For a given model, these effects can be translated into
effects in the cosmic microwave background by
combining them with the fluctuations from the
wall modes and a model for matter such as cold dark matter.
Modifications of this simplified model (wall thickness,
tilt in the final spectrum) should be considered, again to
quantify the sensitivity of the models to
variations in parameters.  Also, so far,
the gravity wave perturbations
have only been calculated for the wall fluctuations
themselves\cite{gar96}.

Along other lines, it would be interesting to see what sorts of
potentials naturally arise in particle theory models.

\section{Acknowledgements}
I thank the Aspen Center for Physics where this work began to take
shape, and the 
Center for Particle Astrophysics and LBNL for hospitality.  
I thank A. Kent, M. Bawendi, P. Ferreira,
J. Garriga, J. Garcia-Bellido, J. Greensite, A. Liddle,
M. Schlieker, N. Sochen, A. Steif
and E. Weinberg for discussions.  I am especially
grateful to A. Anderson, M. Bucher and M. White for numerous
discussions and explanations
and to the latter two in addition for comments on
a draft of this paper.
This work
was supported in part by the Director, Office of
Energy Research, Office of High Energy and Nuclear Physics, Division
of High Energy Physics of the U.S. Department of Energy under Contract
DE-AC03-76SF00098 and in part by the National Science Foundation under
grant PHY-90-21139.

\end{document}